# A Computationally Intelligent Hierarchical Authentication and Key Establishment Framework for Internet of Things

Mohammad Sayad Haghighi, Orwa Nader, Alireza Jolfaei

Our high expectations from Internet of Things (IoT) and how it will positively influence our lifestyles depend on a secure and trusted implementation of it, especially in the sensitive sectors such as health or financial. IoT platforms and solutions must provide Confidentiality, Integrity and Availability (CIA) in a secure and transparent way. Due to the extremely large scales of IoT, traditional centralized solutions for security provisioning cannot be employed in their original form. This article discusses the authentication problem in IoT, which is fundamental to providing CIA. We propose a hierarchical security architecture for this problem and focus on computationally lightweight authentication protocols which can intelligently distribute the computational load across multiple levels and effectively push the load towards upper layers.

## Introduction

Internet of Things (IoT), or as it is sometimes called, Internet of Everything (IoE), has made a technological revolution and is believed to be the next generation of backbone that connects all types of devices including sensors, actuators, GPS devices, mobiles, and almost every other thing [1]. IoT will consist of tens of billions of connected things that can generate, exchange and consume data along with humans. Since IoT is meant to serve in everyday human life scenarios, IoT systems have to be secure and, as far as possible, privacy preserving. However, it is difficult to provide security in heterogeneous networks. It should be taken into considerations that IoT devices are low power and resource-constrained. As such, some of the conventional unintelligent security provisioning techniques, which are often centralized, may not be appropriate and new smart and computationally lightweight solutions must be developed.

In this paper, we focus on the authentication problem as well as key establishment between IoT nodes. We proposed a lightweight and intelligent authentication framework by taking into account the limited processing power of the end nodes. The proposed scheme is hierarchical in order to meet the heterogenous nature of IoT network and is scalable too. In the next section, we numerate IoT ecosystem security challenges and then, present our solution in the subsequent section. We discuss two scenarios for key establishment/authentication and show how we can intelligently reduce our reliance on central third parties to cope with the big scale of IoT network.

## IoT Security Challenges

Some references suggest that IoT architecture should have three layers: Perception, Network and Application [1]–[3]. Perception layer contains sensors, actuators and other physical things, and due to its easy accessibility, is the most vulnerable layer. Since many edge devices have limited processing power, storage or memory capacities, they cannot do complex mathematical computations normally required in traditional security solutions. Therefore, lightweight solutions emerged to address this problem and satisfy the Confidentiality, Integrity and Availability (CIA) requirements in IoT platforms [4]-[6]. For example, some recent research works in the IoT security field focused on lightweight mutual authentication to save integrity of sensors' data in smart cities [7]. Others tried increasing the protection level and focused on preventing physical security violations in public places by using two-factor authentication solutions while preserving privacy [8].

Because of the importance of Machine to Machine (M2M) communication and the role it plays in industrial IoT, reference [9] proposed a lightweight authentication mechanism for M2M communication scenarios based on XOR and hash operations. The simplicity of the operations is obviously due to the weight considerations of the scheme.

IoT security challenges may be divided into three categories [10]: Data Security, Privacy and Trust. In all of these categories, we have to take into our consideration that many IoT tools and equipment have low resources. Many of the issues originate from wireless links, where IoT systems depend on wireless networks which bring along many security issues [3-6][10,11].

In traditional solutions, we rely on trusted third parties like certificate authorities (CA) as trust anchors for key distribution and subsequently, authentication. However, CAs and Public Key Infrastructure (PKI) are built upon asymmetric cryptographic primitives. Asymmetric cryptographic solutions are known to be more complex and resource-demanding than symmetric ones thus are not popular choices for constrained devices. Moreover, the decentralized nature of IoT, rules out the applicability of any central reference as it can turn into a bottleneck in the system [12]. The hierarchical structure of PKI is favorable though as it was designed with the scalability feature in mind.

In the next section, we propose a hierarchical authentication framework that is intelligently designed to accommodate the limitations of IoT network.

# Lightweight Hierarchical IoT Authentication

Our solution employs symmetric lightweight computation techniques. It distributes the authentication computational load intelligently among the nodes across different levels. It has a hierarchical structure and IoT nodes form the leaves of this tree in a security overlay that is made on top of the traditional communication infrastructure. The core idea behind this solution is that parents act as security mediators. They practically implement mediated authentication protocols that are intelligently designed to run on resource-constrained devices.

In this hierarchal architecture, at every level (except the root level), we have one or more head nodes, which work as mediators and have important roles in building authentication paths between the nodes engaged in an authentication or key establishment process.

The proposed hierarchal security architecture has been shown in Fig. 1. This figure shows a four-level architecture though in a broader perspective, it can be extended to have more levels. In the demonstrated figure, District Mediator (DM) is at the topmost level and acts as the root or master head. This could be a district's security anchor. At a lower level, we have a Heads (H) e.g. in every home in the district which govern and cover the security needs of the IoT devices inside houses. Going further down, one gets to the Cluster Head (CH) level. Cluster heads mediate authentication and key establishment requests between two IoT devices in e.g. room. At lowest level lie the IoT nodes (N). As one can imagine, this structure can have more levels and go as far as City Mediator, Country Mediator, Continent Mediator, etc.

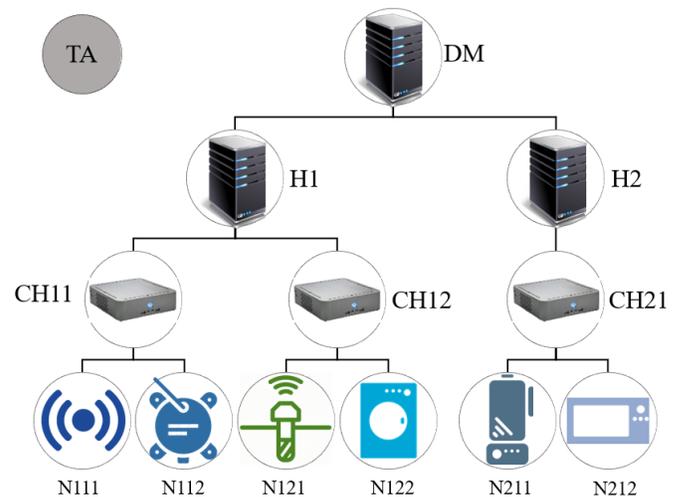

Fig. 1. The proposed hierarchical security overlay architecture for IoT key management and authentication.

In the next subsection, as an example of this architecture application, we show the authentication steps for an in-home key establishment scenario, that is, when the source and destination nodes both lie within the scope of a house head. Then, we generalize the case and take it to the district level in which an IoT object in one house decides to establish a connection with another object in a different house in the same district.

**In-home Authentication**

For in-home scenarios, we argue that a three-level architecture is sufficient, i.e. House Network Head (H), Cluster Head (CH) and IoT Node (N), as shown in Fig. 2. Network Head has sufficient computational power and resources to mediate every in-house connection from the security perspective. It can be the home router for instance that connects the home network to the Internet. Notice that the traffic exchanged between the two IoT nodes does not (necessarily) travel though the H or even CH. As mentioned before, the proposed multi-level solution is a hierarchical overlay security architecture that shows the secure or trusted paths for e.g. authentication and session key establishment. However, the traffic is actually exchanged by the communication plane,

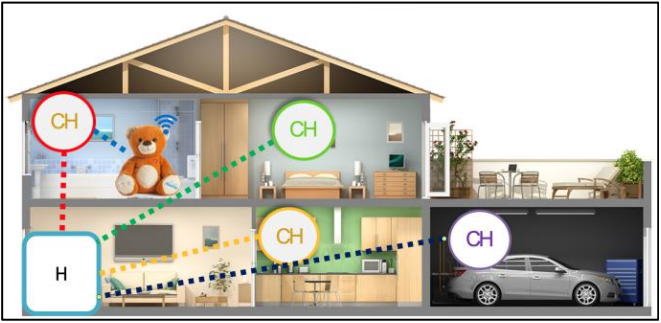

Fig. 2. The key establishment and authentication security overlay architecture for in-house communications.

which can be even direct from one node to another in some circumstances.

Home network is partitioned into different clusters according to its architecture. Every device that has moderate resources, such as a smart Android TV, can be designated as a cluster head. It is important to know that the concept of cluster does not imply being static. For example, in Fig. 2, car itself is a cluster head and all the driver's smart belongings can use the car CH to establish secure connections to the other IoT nodes in his/her home. In that case, the authentication and secure key establishment is mediated by the car CH but once the key is established, the traffic can be routed by the communication plane normally e.g. via 4G/5G or VANET (Vehicular Ad hoc Network).

At the third layer of Fig. 2, there are IoT nodes with the least resources which cannot perform complex computational tasks. In addition to these three layers, we can also have a Trusted Authority (TA) who sets the network up with initial parameters in a setup phase. It can be the nodes' manufacturer in real-world scenarios. In setup phase, TA creates network parameters and publishes it among all network nodes in all layers.

In the installation phase, H establishes secret values with all cluster heads and nodes. CHs and Ns have to register with H over secure channels. In practice, this is done manually. Every CH is registered with H and receives a secret value to communicate with the associated IoT nodes under it in the future. The same process is done for regular IoT nodes. For example, when the house owner buys a new smart IoT device, he/she will manually registers it with the house network head (e.q. by scanning a QR code Head generates) and specifies under which cluster head(s) it will operate. This binding or association of IoT nodes to certain CHs (and CHs to H) by using secret keys is a cryptographic trick introduced in some recent papers [13]. Fig. 3 shows the device installation phase visually, where every CH is registered with H by its ID. This intelligent approach avoids the need for public key/asymmetric cryptography at the edge level. Therefore, low-cost symmetric algorithms can be employed especially for the IoT nodes.

After the installation phase, no (registered) node shall be allowed to get connected to H for getting private parameters in order to prevent impersonation attacks in the future. If a node is to be removed from the system, its registration key with the head can be deleted. This way, the node cannot establish a secure connection anymore.

In the key agreement phase, each IoT node (N) uses the master secret it has been given to establish (symmetric) session keys with the associated CH. These carefully-chosen secret values help both parties authenticate each other during the key establishment phase too.

For N-to-N authentication and (session) key agreement, we use mediators. We can build an authenticated route between nodes in same cluster or in different clusters to help the two nodes authenticate each other. We have three type of authentication: between CHs via H as shown in Fig. 4, between Ns in same cluster via CH and between Ns in different clusters via CHs. Note that since in all the scenarios that an IoT node is involved, it deals with a CH which is more capable in terms of resources,

As we can see in Fig. 4, CH1 and CH2 build an authentic path between themselves with challenge and response mechanism and also establish a session key to have a direct secure channel for future communications. Note that the last message is not (necessarily) sent via H as the key has been established and the two entities can communicate over the communication plane directly.

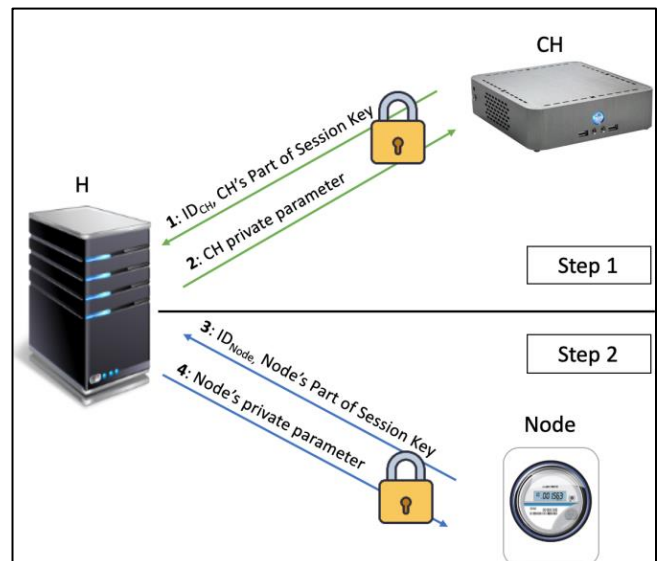

Fig. 3. Device installation or registration phase.

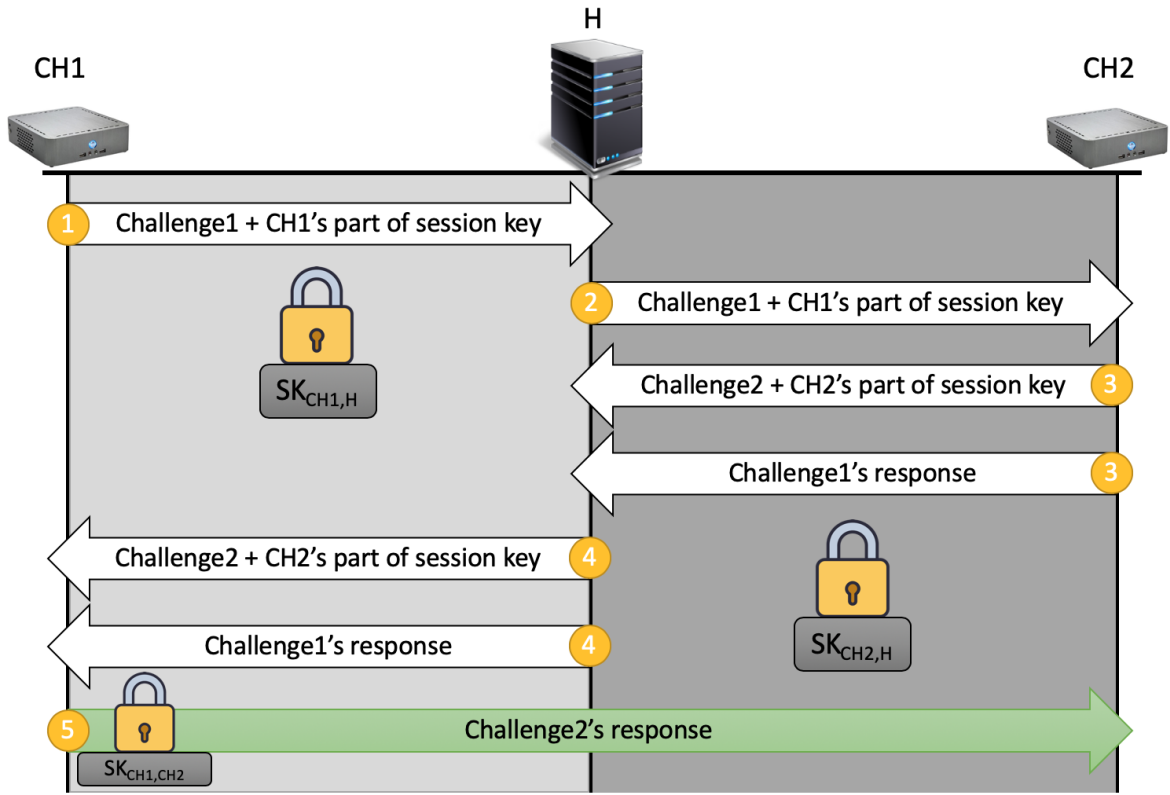

Fig. 4. CH to CH authentication and key agreement mediated by the Head. We have created two secure channels between CH1 and H, and also, between CH2 and H, where all messages are encrypted by a the shared session keys $SK_{CH1,H}$ and $SK_{CH2,H}$, which was built during the registration/installation phase. The last message is sent directly from CH1 to CH2 via the communication plane encrypted with $SK_{CH1,CH2}$ since no more security mediation is required.

What we described was a CH-to-CH authentication and key agreement process. N-to-N authentication and key agreement is very similar with the difference that CHs will mediate the process in that case. If the two nodes reside in different clusters, the authentication shall be mediated by H as the first trusted entity both have access to is the head.

As we noticed, creating authentication path between two sides crossing one mediator, helps in realization of the mutual authentication concept. We suggest using different key parts (seeds) to create session keys each time so that forward secrecy property is maintained.

As was described in the previous scenarios, no IoT node directly engages with another IoT node for authentication or cryptographic purposes. This way the burden of authentication or heavy cryptographic computations can be intelligently pushed towards the more capable device handing the nodes security needs (i.e. CHs in this case) [13]. Once the authentication is done and the mutual session key is established between the two nodes, they can use a lightweight symmetric encryption and message authentication algorithm to directly communicate with each other.

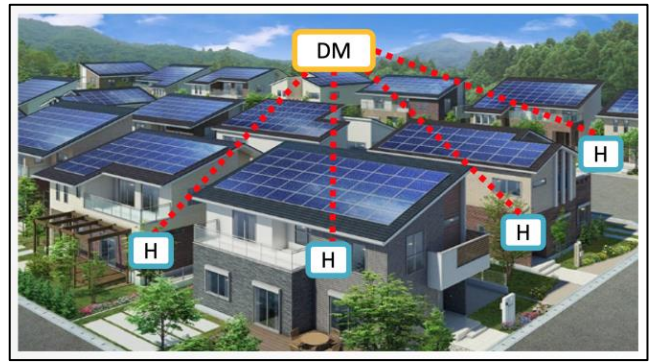

Fig. 5. Visual layout of the District Mediator (DM) in the hierarchical key management and authentication scheme.

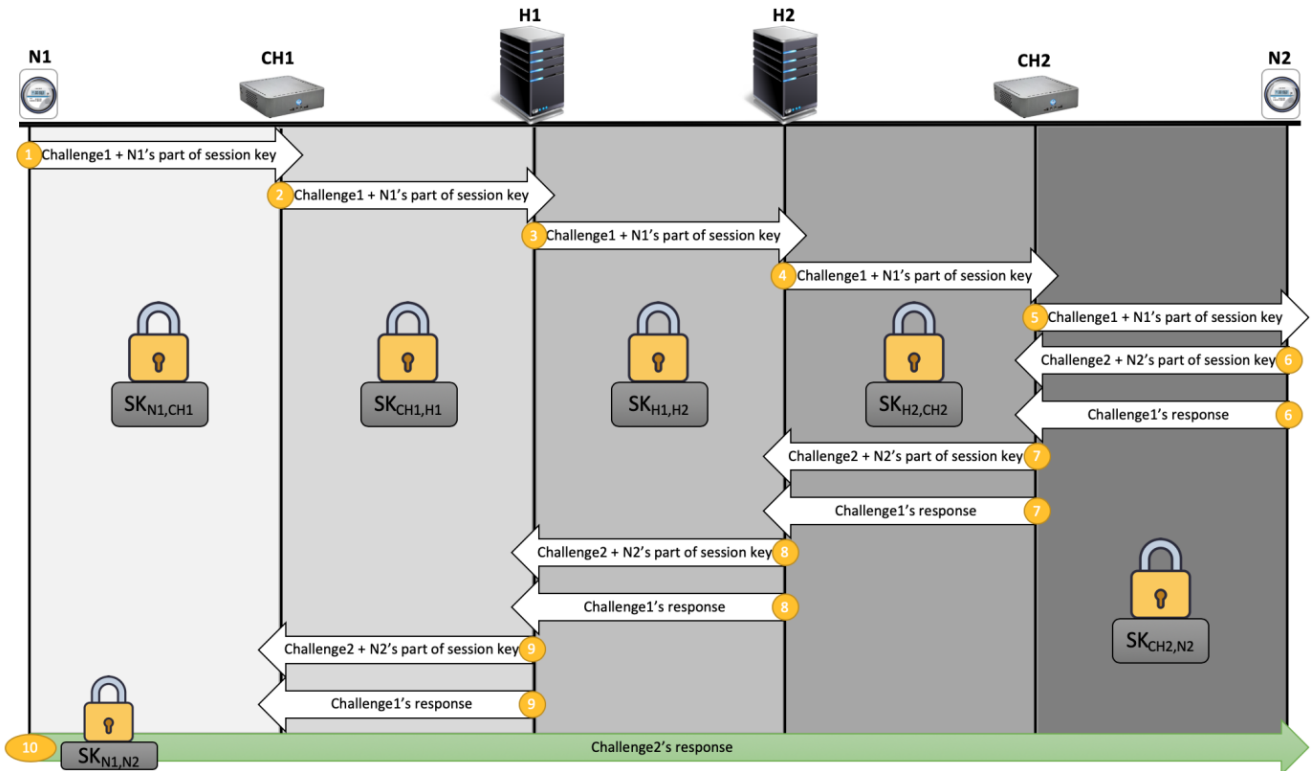

Fig. 6. Building an authentication and key establishment path between two nodes in different houses. The process is mediated by Cluster Heads and Home Heads. We have five secure channels between the nodes and mediators along the path. The last message is sent directly via the communication plane and encrypted with $SK_{N1,N2}$ as from this point on, no help in security mediation is required.

**Generalization to Multi-level Authentication**

The scenario we discussed in the previous subsection was an in-home one. Now, we show how this idea can be scaled up to support bigger IoT networks. Assume that we want to add one more layer to the hierarchy to make district-wide IoT networks in which every node in that district can establish secure connection to any other node. More levels like city, country or continent can be similarly built up.

For the district level, we need to need another object to the previous model, which acts as the mediator for in-home's heads. We call this new player District Master (DM) as shown in Fig. 5. In the installation phase, we have to add/register Hs to DM and give each H an association key.

Now, if one node from one house wants to connect to another node in a different house, for the sake of authentication and key establishment, it shall go through its CH, H and DM to get to the other side's H, CH and finally N. Fig. 6 shows the authentication and key establishment process between two such nodes associated with two different houses. Where, the Heads have made a symmetric key by the help of District Master before. In addition, all mediators pass the key establishment requests/response to the next mediator, and have to decrypt and re-encrypt messages and ensure mutual authentication in every link.

Note that like before, the built path is only used for authentication and key establishment. The traffic exchanged between these nodes can flow through the communication plane normally. Moreover, this idea does not lock any IoT object down to the house perimeter. Consider a house owner is driving and its smart phone wants to fetch some information from a neighbor's sensor. If such a permission has been granted, its phone negotiates with the car CH and then with H and DM to get to the destination sensor's H and CH. After the destination node is authenticated and a key is established via the mediators, they can talk directly via the communication plane, even if the driver is on the move.

# Conclusion

The important role of IoT in our lives and the its spread across many segments has raised concerns in areas that may contain sensitive prescriptive systems e.g. health and finance. Security of such interconnected systems has now become very crucial but to make any prescription for such a network, one has to take into consideration the limited processing power of IoT equipment and gadgets. In this paper, we proposed a lightweight authentication and key management framework with hierarchical structure that intelligently pushes the load of security services such as authentication towards mediators who are more capable or have more resources. One can take this framework and develop a protocol to realize a practical instance of it.

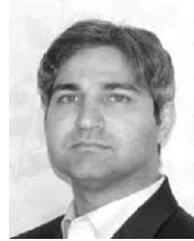
**Mohammad Sayad Haghighi** is the Head of IT Department at the School of Electrical and Computer Engineering, University of Tehran, Iran. Prior to joining the University of Tehran, he was an Assistant Professor at Iran Telecom Research Center. He has done a postdoctoral study at Deakin University during 2012 and 2013. His research interests are cryptography, cyber security and wireless ad hoc networks. Dr. Sayad has won several national grants including some from Iran National Science Foundation (INSF). He is the director of ANSLab (Advanced Networking and Security research Laboratory) and a Senior Member of IEEE.

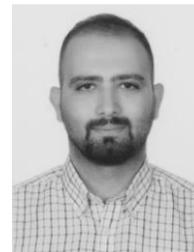
**Orwa Nader** received Eng. in Networks and Information Systems from Al-baath University, Syria in 2010, and his first M.Sc. in Information and Decision Support Systems in 2016 from Higher Institute of Applied science and Technology, Syria, and his second M.Sc. in Information Technology from the University of Tehran, Iran in 2018. He is currently a Ph.D. candidate in Information Technology at the University of Tehran. His research interests include cyber security as well as security protocols especially in the context of Internet of Things.

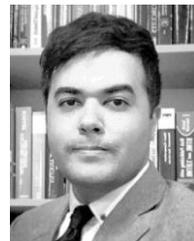
**Alireza Jolfaei** received a Ph.D. degree in Applied Cryptography from Griffith University, Australia. He is an Assistant Professor in Cyber Security at Macquarie University, Australia. Prior to this appointment, he worked as an Assistant Professor at Federation University, Australia and Temple University in Philadelphia, USA. His current research areas include Cyber Security, IoT Security, Human-in-the-Loop CPS Security, Cryptography, AI and Machine Learning for Cyber Security. He has authored over 50 peer-reviewed articles on topics related to cyber security. He is a Senior Member of the IEEE and an ACM Distinguished Speaker on the topic of Cyber-Physical Systems Security.